\documentclass[11pt]{iopart}

\usepackage{graphicx}
\usepackage{iopams}
\usepackage{amssymb}

  \expandafter\let\csname equation*\endcsname\relax
  \expandafter\let\csname endequation*\endcsname\relax
  
\usepackage{amsmath}
\usepackage{bm}
\usepackage{color}
\usepackage[english]{babel}

\newcommand{\ket}[1]{{|{#1}\!\!>}}
\newcommand{\bra}[1]{{<\!\!{#1}|}}

\newcommand{\spi}[2]{{\left[ \begin{array}{c} {#1} \\ {#2} \end{array} \right]}}

\begin{document}

\title{Natural Parameterization of Two-Qubit States}

\author{K B Wharton}

 \address{San Jos\'e State University, Department of Physics and Astronomy, San Jos\'e, CA 95192-0106}

\begin{abstract}
Any pure two-qubit state can be represented by six real angles, with a natural parameterization indicated by the bipartite structure.  After explicitly identifying all of these angles for the first time, it is found that the parameters can always be completely separated into two ``dynamically local'' spinor components.  Specifically, given local Hamiltonians at the locations of the two qubits, unitary dynamics on each spinor can be implemented separately without losing any entanglement information in the full state.  Such a conclusion also follows from a phase-fixed version of the Schmidt decomposition.

\end{abstract}


\setlength{\baselineskip}{1.1\baselineskip} 

\section{Introduction}

Pure two-qubit states reside in the 4-dimensional Hilbert space $\mathcal{H}_4$, and they are of great significance to a wide range of fields from quantum computation to quantum foundations.  When constrained only by an overall normalization, the space of all such states corresponds to the geometry a 7-sphere, $\mathcal{S}^7$.  Typically, one ignores a global phase, leaving six (real) angles that parameterize the space of all pure two-qubit states.  (Although ignoring a single phase angle cannot be trivially accomplished without introducing topological phase-jumps and also breaking the $\mathcal{S}^7$ symmetry; for a discussion of the analogous situation in $\mathcal{H}_2$, see \cite{WhartonKoch}.)

A natural parameterization of this space is motivated whenever $\mathcal{H}_4$ results from a bipartite system, of one qubit each.  This is not to imply a restricted attention to the special separable states, where each qubit lies in $\mathcal{H}_2$ .  Even for entangled two-qubit states, perhaps encoded in the spins of spatially-separated fermions, it is known how all locally-measurable parameters can be extracted from the full $\mathcal{H}_4$.  This is accomplished by taking partial traces of the full 4x4 density matrix to produce two 2x2 density matrices, one for each qubit.  These partial traces will generally correspond to mixed states, and can each be mapped onto a vector in the Bloch ball.  

Leaving aside the case of maximally-entangled states (for now), the orientation of these Bloch ball vectors define two natural angles for each qubit.  And with four angle parameters already ``chosen'' by the bipartite structure, there should be exactly two remaining angles that specify the full entangled state.  Remarkably, one of these angles has never been explicitly identified.

One of the two remaining angles is directly related to the ``concurrence'', a well-known parameter \cite{Wootters1,Wootters2}.  The concurrence is a measure of entanglement, and is evident in the partial traces, as it is also related to the length of both Bloch ball vectors.  But while it is encoded in each partial trace, no unitary operation on a single qubit can alter this parameter.  (Such operations will be termed ``local'' throughout this paper, in the sense that they can be performed on one of the two qubits, even without access to the other.)

The sixth natural angle parameter needed to fully characterize states in $\mathcal{H}_4$ is developed below; this new angle is referred to as the ``recurrence''.  Unlike the concurrence, the recurrence is \textit{not} evident from the individual partial traces.  It is also unlike the concurrence in that it \textit{can} be altered via a local operation.  This curious combination might make it seem more ``nonlocal'' than the concurrence, but the below analysis will demonstrate that such a conclusion is almost certainly not warranted.

In 2001, a few years after the concurrence was identified \cite{Wootters1,Wootters2}, Ku\'s and \.Zyczkowski nearly completed the parameterization of pure two qubit states \cite{Kus}.  There they identified the 3D manifold of maximally entangled states, the 4D manifold of separable states, and (for any given concurrence) 5D manifolds of partially entangled states.  But the 5D manifolds were not formally parameterized in that work, and that same year, another key paper steered such research in a different direction.  

That paper, due to Mosseri and Dandoloff \cite{Mosseri}, demonstrated that a seemingly natural Hopf fibration of the $\mathcal{S}^7$ space of pure $\mathcal{H}_4$ states was sensitive to entanglement (and therefore concurrence).  Much of the subsequent analysis concerning parameterization of such states used this observation as a starting point.  But this procedure formally breaks the symmetry between the two qubits: one of the two qubits has to be mapped to the base space of the Hopf fibration, and the other one is not.  In the special case of separable states, the second qubit can be found in the fiber, but for any non-separable state even this ``symmetry'' vanishes \cite{Mosseri,Levay}, and the natural bipartite structure is lost.  For example, in one recent attempt at a full parameterization \cite{Wie}, this broken symmetry between the two qubits arguably contributes to problematic coordinate singularities and phase discontinuities.

The present work returns to the Ku\'s-\.Zyczkowski framework and completes the parameterization of partially entangled states in terms of parameters that are as ``local'' as possible, in that they are identifiable with the partial trace of one qubit or the other.  As noted above, only one new angle is needed. Section 3 then turns to an analysis of this new angle, the recurrence, and reveals that it can be treated as the sum of two ``local'' phase angles, an apparent explanation of its curious features.  Section 4 connects these result to the Schmidt decomposition in order to resolve the maximally-entangled limit, and also distinguishes the recurrence from a similar but ill-defined ``Schmidt angle'' that has appeared in the literature \cite{Sjoqvist,Loredo}.  A seemingly novel practical application is then detailed in section 5: the ability to exactly transform entangled states in $\mathcal{H}_4$ via two independent transformations on $\mathcal{H}_2$. 

Some foundational issues are raised by this analysis.  The final section touches on some of these questions, concerning nonlocality, the meaning of quantum phases, and possible future extensions of this work.  While there is no obvious path to extending this analysis to mixed- or multi-partite states, these results nevertheless provide a useful new perspective on the simplest nontrivial bipartite quantum system.

\section{Six-Angle Parameterization}

In a fixed basis, the most general pure two-qubit state can always be written in terms of four complex parameters $(a,b,c,d)$:
\begin{equation}
\label{eq:psidef}
\ket{\psi}=a \ket{00} + b \ket {01} + c \ket {10} + d \ket {11}.
\end{equation}
An overall normalization, $|a|^2+|b|^2+|c|^2+|d|^2=1$ will be assumed and enforced throughout.  The concurrence of this state is defined as $\mathcal{C}=2|ad-bc|$; this goes to zero for a separable state and to unity for a maximally-entangled state \cite{Wootters1,Wootters2}.  The concurrence angle, $0\le\chi\le\pi/2$, can then be defined as $\mathcal{C}=\sin{\chi}$. 

The individual properties of each of the two qubits can be found by taking partial traces of $\ket{\psi}\bra{\psi}$.  For the first qubit (\textit{i.e.} the ``0'' in $\ket{01}$), this yields
\begin{equation}
\label{eq:rho1}
\rho^{(1)}= \begin{pmatrix}
|a|^2+|b|^2& ac^*+bd^* \\
a^*c+b^*d& |c|^2+|d|^2
\end{pmatrix}.
\end{equation}
The first qubit's Bloch ball vector $\bm{n}^{(1)}$ can then be determined from $\rho^{(1)}=(\bm{I}+\bm{n}^{(1)}\cdot \bm{\sigma})/2$, where $\bm{\sigma}$ is the usual vector of Pauli matrices and $\bm{I}$ is the identity.  The components of this vector are given by
\begin{align}
\label{eq:n1}
|a|^2+|b|^2-|c|^2-|d|^2&=n^{(1)}_z\\
2(ac^*+bd^*)&=n^{(1)}_x-in^{(1)}_y. \nonumber
\end{align}

Note the magnitude of $\bm{n}^{(1)}$  is always exactly $\cos{\chi}$.  If its magnitude is non-zero, its direction in spherical coordinates $(\theta_1,\phi_1)$ can be found from $\cos(\theta_1)=n^{(1)}_z/ \cos(\chi)$ and $\phi_1=-\arg(ac^*+bd^*)$.  The same procedure can be run for the second qubit, yielding another Bloch ball vector $\bm{n}^{(2)}$ with parameters
\begin{align}
|a|^2+|c|^2-|b|^2-|d|^2&=n^{(2)}_z \\
2(ab^*+cd^*)&=n^{(2)}_x-in^{(2)}_y. \nonumber
\end{align}
This vector also has magnitude $\cos{\chi}$, and defines two more angles $(\theta_2,\phi_2)$ for non-zero $\bm{n}^{(2)}$.  These angles become undefined for the special case of maximally entangled states ($\chi=\pi/2$), as that is when $\bm{n}^{(1)}=\bm{n}^{(2)}=0$, but this problem will be set aside until Section 4.  Until then, we will assume $\chi\ne\pi/2$. 

The primary result of this paper is that the four complex parameters in (\ref{eq:psidef}) can be written in terms of these five angles $(\chi,\theta_1,\phi_1,\theta_2,\phi_2)$ and one additional angle $\gamma$:
\begin{align}
\label{eq:abcd}
a=\left[ \cos\frac{\chi}{2}\cos\frac{\theta_1}{2}\cos\frac{\theta_2}{2}e^{i\gamma/2}+\sin\frac{\chi}{2}\sin\frac{\theta_1}{2}\sin\frac{\theta_2}{2}e^{-i\gamma/2} \right] e^{-i(\phi_1+\phi_2)/2}, \\
b=\left[ \cos\frac{\chi}{2}\cos\frac{\theta_1}{2}\sin\frac{\theta_2}{2}e^{i\gamma/2}-\sin\frac{\chi}{2}\sin\frac{\theta_1}{2}\cos\frac{\theta_2}{2}e^{-i\gamma/2} \right] e^{-i(\phi_1-\phi_2)/2}, \nonumber \\
c=\left[ \cos\frac{\chi}{2}\sin\frac{\theta_1}{2}\cos\frac{\theta_2}{2}e^{i\gamma/2}-\sin\frac{\chi}{2}\cos\frac{\theta_1}{2}\sin\frac{\theta_2}{2}e^{-i\gamma/2} \right] e^{+i(\phi_1-\phi_2)/2}, \nonumber \\
d=\left[ \cos\frac{\chi}{2}\sin\frac{\theta_1}{2}\sin\frac{\theta_2}{2}e^{i\gamma/2}+\sin\frac{\chi}{2}\cos\frac{\theta_1}{2}\cos\frac{\theta_2}{2}e^{-i\gamma/2} \right] e^{+i(\phi_1+\phi_2)/2}. \nonumber
\end{align}
This natural parameterization defines a new angle $\gamma$, hereafter referred to as the ``recurrence''.  It can be checked by the tedious process of recovering the five known angles using their well-defined relationships with $(a,b,c,d)$; in all five cases, $\gamma$ cancels exactly.  Also, note that these expressions automatically obey overall normalization, and that the global phase has been chosen such that the complex concurrence $2(ad-bc)$ is always real.

The known 5 angles in the above equations can easily be written in terms of $(a,b,c,d)$, so in principle one can invert the above equations to find $\gamma$ as a function of $(a,b,c,d)$.  Unfortunately, direct substitution of all five angles leads very complicated expressions.  One simpler way to extract $\gamma$ turns out to be:

\begin{equation}
\label{eq:gamma1}
\sin(\gamma)= \frac{2 \, Im(ad+bc)}{\cos \chi \sin \theta_1 \, \sin \theta_2}.
\end{equation}

Contrary to appearances, this expression does not depend on the global phase, because the global phase has already been fixed to make $(ad-bc)$ real.\footnote{Alternatively, $Im(ad+bc)$ could be read as the magnitude of the component of $(ad+bc)$ perpendicular to $(ad-bc)$ in the complex plane, and this is evidently independent of global phase.}  Still, this expression is ambiguous for separable states when $(ad-bc)=0$, because then the phase cannot be fixed.  It also fails when either $\theta_1$ or $\theta_2$ goes to $0$ or $\pi$, and also for maximally entangled states.  Furthermore, it is hard to read much physical significance from even this simple form of (\ref{eq:gamma1}); the only obvious implication is that under a particle-exchange (swapping $b$ and $c$), $\gamma$ is unchanged.

Still, the failures of (\ref{eq:gamma1}) are perfectly explicable; maximally entangled states have already been set aside, and for separable states ($\chi=0$) $\gamma$ is indistinguishable from a global phase, evident from (\ref{eq:abcd}).  When either $\theta$ goes to $0$ or $\pi$, the corresponding Bloch vector is on the z-axis, making $\phi$ undefined.  It should be evident from (\ref{eq:abcd}) that $\gamma$ necessarily inherits this familiar coordinate singularity of $\phi$ for such states.  The solution to this latter problem, at least, seems to be a more careful treatment of the original Bloch-ball geometry.  The next section will demonstrate that such a procedure leads to a natural interpretation of $\gamma$.

\section{Interpretation of the Recurrence}

Before resolving the coordinate singularities from the previous section, it is useful to understand how the recurrence $\gamma$ can be varied.  It turns out that a local interaction at the location of \textit{either} qubit can change $\gamma$, via a rotation of the local qubit around its own Bloch-ball vector (as determined by the partial trace).  This interaction clearly does not change the local partial trace, and of course it cannot change the distant partial trace, but it does usually change the full state.  Specifically, such an interaction linearly affects the recurrence, while leaving all other angles constant. (For a brief proof of these claims, see the Appendix.)

If one thinks of the recurrence as a global property of the entangled state, residing nowhere in spacetime, such local-based manipulations of $\gamma$ might almost seem like a non-measurement form of nonlocality.  These manipulations have nothing to do with measurements (they are perfectly unitary), and yet somehow experimenters at the locations of the two qubits both have independent control of this single, locally-unmeasureable parameter.  Any useful interpretation of this angle should resolve this mystery one way or the other.

Turning back to the coordinate singularity that occurs when either partial trace lies on the z-axis, one obvious approach is to replace each pair of angles with the cartesian components of a unit 3-vector, as the latter form does not suffer from any special poles.  However, it is far from simple to rewrite (\ref{eq:abcd}) in terms of these components, because the angles that appear in (\ref{eq:abcd}) are all \textit{half}-angles.  As half-angles are familiar from spinor notation, one promising option is to take the Bloch-ball vectors from the two partial traces and construct two corresponding ``local spinors'' (renormalized, or effectively projected out onto the surface of the Bloch sphere):

\begin{align}
\label{eq:4vec}
\ket{\phi_1} = e^{i\alpha_1/2}\spi{\cos  \frac{\theta_1}{2} e^{-i \phi_1/2}}{\sin  \frac{\theta_1}{2} e^{+i \phi_1/2}} \equiv \spi{A}{B}\\
\ket{\phi_2} = e^{i\alpha_2/2}\spi{\cos  \frac{\theta_2}{2} e^{-i \phi_2/2}}{\sin  \frac{\theta_2}{2} e^{+i \phi_2/2}}\equiv \spi{C}{D}\nonumber 
\end{align}

Here the angle representations have the same coordinate singularities as (\ref{eq:abcd}), but the complex numbers $(A,B,C,D)$ have no such problems.  (For example, the real and imaginary parts of $A$ and $B$ comprise a unit 4-vector on $\mathcal{S}^3$, without special poles.)  Also note that these expressions introduce local phases, $\alpha_1$ and $\alpha_2$, associated with one qubit or the other.  One cannot cleanly remove these phases without reintroducing coordinate singularities in the spinors. \cite{WhartonKoch}

Defining two new phase angles (on top of the six angles above) may seem excessive, given that one cannot hope to use eight independent angles to parameterize a seven-sphere.  However, this problem will be neatly resolved below when some of these angles are found to be related.

In terms of these ``local'' parameters, for now ignoring the $\alpha$'s and the $\gamma$, the other terms in (\ref{eq:abcd}) can be neatly expressed as
\begin{align}
\label{eq:abcd2}
a&=AC\cos\frac{\chi}{2} + B^*D^*\sin\frac{\chi}{2} \\
b&=AD\cos\frac{\chi}{2} -B^*C^*\sin\frac{\chi}{2}  \nonumber \\
c&=BC\cos\frac{\chi}{2} -A^*D^*\sin\frac{\chi}{2}  \nonumber \\
d&=BD\cos\frac{\chi}{2} +A^*C^*\sin\frac{\chi}{2}. \nonumber 
\end{align}
Substituting (\ref{eq:4vec}) into (\ref{eq:abcd2}), one finds that (\ref{eq:abcd}) is recovered \textit{exactly} under the simple assignment
\begin{equation}
\label{eq:money}
\alpha_1+\alpha_2=\gamma.
\end{equation}

This result resolves the mysterious local control of the nonlocal recurrence.  Instead of thinking of the recurrence as a single angle, and the global phase as another single angle, it is far more natural to eliminate both of these in favor of $\alpha_1$ and $\alpha_2$.  These latter angles are clearly local, being associated with one particular qubit, and their sum is the global recurrence.  This explains the local control of the recurrence; an interaction at either qubit can change the local phase $\alpha$, and therefore can change a global $\gamma$.

The conclusion is that given the bipartite structure, the most natural parameterization of two-qubit states on $\mathcal{S}^7$ is in terms of two local spinors ($\ket{\phi_1}$ and $\ket{\phi_2}$, each on $\mathcal{S}^3$), and one shared angle, the concurrence ($\chi$).  No local unitary interaction can change $\chi$, so this parameter is just a global constant (so long as the qubits are separated).  A potential resolution for the problematic limit of maximally-entangled states will be proposed in the next section.

\section{Connection to Schmidt Decomposition}

Given a value of the concurrence $\chi$, (\ref{eq:abcd2}) indicates the precise relationship between the full state $\ket{\psi}$ (in $\mathcal{H}_4$) and the two local normalized spinors $\ket{\phi_1}$ and $\ket{\phi_2}$ (each in $\mathcal{H}_2$).  If $\mathcal{P}$ is defined as a particular parity-inversion transformation that takes a normalized spinor to its antipodal point on the Bloch sphere,
\begin{equation}
\label{eq:parity}
\mathcal{P} \spi{A}{B} = \spi{B^*}{-A^*},
\end{equation}
then it follows from (\ref{eq:abcd2}) that any entangled state can be neatly written as a function of these two local spinors and the concurrence angle:
\begin{equation}
\label{eq:key}
\ket{\psi} = \left(\cos\frac{\chi}{2}\right) \ket{\phi_1} \otimes \ket{\phi_2} + \left(\sin\frac{\chi}{2}\right) \mathcal{P} \ket{\phi_1} \otimes \mathcal{P} \ket{\phi_2}.
\end{equation}
This is evidently a Schmidt decomposition of the original state, with no phase ambiguities.  (The phase relationship between $\ket{\phi}$ and $\mathcal{P}\ket{\phi}$ is fixed by the definition (\ref{eq:parity}), and the phase of $\ket{\psi}$ is fixed by the earlier requirement that $(ad-bc)$ is real.)

This form is distinct from the standard Schmidt decomposition, which does generally have phase ambiguities that are normally absorbed into the Schmidt basis itself.  When a phase is pulled out of the Schmidt basis, for example as in \cite{Sjoqvist}, the resulting ``Schmidt angle'' can look suspiciously like the recurrence.  However, there are several key differences.  The first is that there is no absolute definition of that Schmidt angle; only its relative angles are meaningful.  The recurrence, on the other hand, can be defined for almost any partially-entangled state via (\ref{eq:gamma1}).  

A more important difference is that without a well-defined phase relationship between the two orthogonal spinors corresponding to $\ket{\phi}$ and $\mathcal{P}\ket{\phi}$, it is impossible to have a well-defined evolution of the Schmidt angle as the Schmidt basis changes.  Indicative of this problem is that the Schmidt angle can be held fixed under any local evolution of a single qubit, as seen in \cite{Sjoqvist}.  The recurrence, however, generally changes under such transformations, as in the Appendix.

Turning back to the phase-fixed Schmidt decomposition (\ref{eq:key}), note that this expression continues to be valid for \textit{maximally} entangled states.  The local-spinor framework developed above must therefore also continue to be usable.  The earlier problem was that the partial traces go to zero, along both local Bloch ball vectors $\bm{n}^{(1)}$ and $\bm{n}^{(2)}$, which loses the directionality for any outward projection onto the surface of the Bloch sphere.  But the expression (\ref{eq:key}) is not in terms of $\bm{n}$, it is in terms of the local normalized spinors $\ket{\phi_1}$ and $\ket{\phi_2}$.  If one takes these spinors to be the relevant local entities, no problems occur in the maximally entangled limit -- although the local spinors do become underdetermined.  

Consider the following illustrative example.  According to (\ref{eq:key}), the (maximally entangled) singlet state $\ket{\psi}\!\!=\!\!(\ket{01}-\ket{10})/\sqrt{2}$ can be represented in a variety of ways.  This singlet state corresponds to $\chi=\pi/2$ and any pair of local spinors for which $\ket{\phi_2}\!\!=\!-\mathcal{P}\ket{\phi_1}$.   In other words, even if $\bm{n}^{(1)}=0$, at the very center of the Bloch ball, one can form a useful $\ket{\phi_1}$ by projecting onto the surface of the Bloch sphere in \textit{any} direction.  The choice of direction for one qubit fixes the choice for the other (in this case, they are always opposite), and the relative phase is fixed as well.  No matter what direction was initially chosen, one can use the resulting local-spinors to represent the full state, using (\ref{eq:key}).  Note this freedom of choice is only available at the precise value $\chi=\pi/2$.

For this example, one way to represent the singlet state is $\ket{\phi_1}={1\choose 0}$, and $\ket{\phi_2}={0\choose 1}$.  Notice that the local spinor phases matter a great deal: changing the phase of $\ket{\phi_1}$ changes the full state.  If $\ket{\phi_1}$ is rotated around its own axis (as per the Appendix), this changes $\alpha_1$.  When $\alpha_1$ has increased by $\pi$, $\ket{\phi_1}$ picks up a factor of $i$.  Inserting $\ket{\phi_1}={i\choose 0}$ back into (\ref{eq:key}) one finds $\ket{\psi}$ has rotated into $i(\ket{01}+\ket{10})/\sqrt{2}$, and is no longer in a singlet state.  This is exactly the correct transformation of a singlet state if one qubit's Bloch vector undergoes a $\pi$ rotation around the z-axis.  

The local-spinor picture developed in the previous section is therefore still usable for maximally-entangled states, if one uses (\ref{eq:key}) to represent the full state.  There are subtle problems with this resolution, but they only come up when $\chi$ can be changed.  (In this case, the transitions to and from $\chi\!=\!\pi/2$ looks awkward; this issue will be addressed in a future publication.)    But if all operations on the two qubit state are local to one qubit or the other, then the ``local spinor'' model from the previous section works perfectly well.  This result points to a interesting application: the ability to implement exact dynamical evolution on the separate local spinors, without losing any information, even if the full state is entangled.  Such an application will be developed in the next section.

\section{Application: Separable Dynamics for Entangled States}

When manipulating two-qubit states corresponding to a physically-separated system (say, two spin-1/2 particles at well-defined locations $A$ and $B$), it is thought to be generally necessary to treat the two-qubit system as a single entity in $\mathcal{H}_4$.  But for separable states, at least, it is natural to envision exclusively local parameters residing at $A$ and $B$, each in $\mathcal{H}_2$.  The benefit is not merely that the local parameters encode possible local measurements, but also that any local unitary interaction can be implemented using operations on $\mathcal{H}_2$ rather than $\mathcal{H}_4$.  Such a simplification has proven to be too useful to ignore.

The above results indicate that most of these same simplifications are available for entangled states.  Of course, the correlations between distant measurements can only be explained in terms of the full state in $\mathcal{H}_4$.  But setting non-unitary measurements aside, local unitary transformations can be implemented without any operations on $\mathcal{H}_4$.  The key point is that knowledge of the local phases $\alpha_1$ and $\alpha_2$ for each individual qubit can be used to reconstruct the recurrence via (\ref{eq:money}).  Since no local unitary transformation can change the concurrence, all six angles that define $\ket{\psi}$ can be recovered from the local spinors, and the dynamics becomes fully separable -- even for entangled states.

The only practical barrier to implementing these separable dynamics is a small complication concerning the most general form of a local Hamiltonian $\bm{H}$ experienced by a single qubit (at either $A$ or $B$):
\begin{equation}
\label{eq:Hdef}
\bm{H}=H_I \bm{I} + \bm{v}\cdot\bm{\sigma} = \bm{H_I}+\bm{H_\sigma}.
\end{equation}
Here $\bm{v}$ is three-vector, and $\bm{H_\sigma}=\bm{v}\cdot\bm{\sigma}$ is the part of the Hamiltonian that can rotate the state on the Bloch sphere (say, an interaction with a magnetic field).  $\bm{H_I}$ is the spin-independent part of the energy (say, a rest mass or gravitational potential).  

So long as $\bm{H_I}=0$, there are no complications.  It is easy to ascertain that any unitary evolution by $\bm{H_\sigma}$, on either or both qubits, cannot alter the phase condition assumed by (\ref{eq:abcd}): if $(ad-bc)$ is real at one time, it will always be real.  Furthermore, it can also be seen that $[\exp(i\bm{H_\sigma}),\mathcal{P}]=0$.  Since these operations commute, the presence of $\mathcal{P}$ in (\ref{eq:key}) does not pose any difficulty for unitary operations based on $\bm{H_\sigma}$.  For example, given the initial local-spinor representation $\ket{\phi_1}$ and $\ket{\phi_2}$, a local unitary operation on the first qubit simply leads to an updated spinor,
\begin{equation}
\label{eq:ex1}
\ket{\phi'_1}=\exp\left({\frac{-i\bm{H_\sigma}t}{\hbar}}\right)\ket{\phi_1},
\end{equation}
which can be used in (\ref{eq:key}) to recover the full entangled state after this operation.

The complication is when $\bm{H_I}\ne0$.  The corresponding unitary operation \textit{will} add a complex component to $(ad-bc)$, and does \textit{not} commute with $\mathcal{P}$.  Fortunately, these problems exactly cancel out; after all, it is simple to see that applying $\bm{H_I}\otimes\bm{I}$ to the full entangled state will only lead to a global phase change of $\ket{\psi}$.  The obvious solution, then, is just to manually set $\bm{H_I}=0$.  (Another option, that does not lose this global phase information, is entertained in the final section.)

Once the identity-portion $\bm{H_I}$ of each local Hamiltonian is set to zero, the dynamics becomes fully separable, in the following sense.  One can take the initial entangled state, and separate it into its local spinor components $\ket{\phi_1}$ and $\ket{\phi_2}$ as described in section 3.  (One should also calculate the concurrence angle, $\chi$; this will be a constant throughout.)  Each of these local spinors then can be evolved \textit{locally}, via unitary operations on $\mathcal{H}_2$ encountered by the separate qubits.  Care must be taken to retain the proper phase, so the typical single-qubit-gate terminology might be problematic, but this issue is addressable.    Finally, one can take the evolved-spinors $\ket{\phi'_1}$ and $\ket{\phi'_2}$ and use (\ref{eq:key}) to reproduce the entire entangled state.  Given that  any such local operations commute with $\mathcal{P}$, success of this protocol should be evident from the phase-fixed form of the Schmidt decomposition in (\ref{eq:key}).

Using this technique, many of the simplifications enjoyed for separable states can also be utilized for entangled states.  Namely, if the bipartite state is physically separated, all of the dynamics can be found in terms of single-spinor transformations, without any operations on $\mathcal{H}_4$.  It is crucial to keep in mind that $\ket{\phi_1}$ is not a standard quantum state representing the first qubit, but is instead a spinor that (together with $\chi$) encodes everything about the state at that location (how it interacts with any local Hamiltonian, and what local measurements will find).  In particular, one would have to use a variant of the Born rule to extract probabilities.  For instance, when measuring the first qubit, (\ref{eq:key}) indicates that the probability of a spin-direction measurement corresponding to a local eigenstate $\ket{\psi_1}$ would be
\begin{eqnarray}
\label{eq:Born2}
P(\psi_1) &=& \cos^2\frac{\chi}{2} |\!\!<\!\!\psi_1\!\ket{\phi_1}|^2+\sin^2\frac{\chi}{2} |\!\!<\!\!\psi_1\!|\mathcal{P}\ket{\phi_1}|^2 \\
&=& \cos{\chi} |\!\!<\!\!\psi_1\!\ket{\phi_1}|^2 + \sin^2\frac{\chi}{2}. \nonumber 
\end{eqnarray}
For separable states, one recovers the ordinary Born rule; for maximally entangled states, this probability is always $1/2$. 

\section{Discussion}

Parsing the space of pure two-qubit states into the natural 6-angle parameterization of (\ref{eq:abcd}) has led to several interesting consequences that seem not to have been obvious from the Schmidt decomposition alone.  Section 3 demonstrated that instead of thinking of the recurrence angle as a global, unlocalized parameter, it is arguably more natural to split it into two local phase angles, $\alpha_1$ and $\alpha_2$.  This general ``parameter localization'' of the two qubit state is surprising, given the inherent non-local correlations observed when such states are actually measured.

Even more striking, the analysis from section 5 indicates these local parameters can be used to reproduce the exact dynamics of the full entangled state, so long as the interactions on each individual qubit act like local unitary transformations.  This result might have arguably been evident from the Schmidt decomposition, but a literature search has so far failed to find any analogous observation.  More likely, without the phase-fixed version in (\ref{eq:key}), it has been unclear how to properly transform the phases under local dynamical evolution.  The confounding influence of the spin-independent portion of the local Hamiltonian, $\bm{H_I}$, may have also obscured the above result. 

There are clearly potential cases where this result could lead to a practical computational speed-up when simulating two-qubit dynamics.  If both entangled qubits are subjected to different local Hamiltonians, the standard procedure is to take both Hamiltonians (and both qubits) into account for the entirety of the evolution, via operations on $\mathcal{H}_4$.  But using the above results, one could instead treat the two qubits separately, via single-qubit transformations on $\mathcal{H}_2$ (keeping track of the local phases).  Instead of having to recalculate the global behavior each time one of the two Hamiltonians changed, one need only recompute the local spinor corresponding to that single qubit.  After evolution, the exact full state can always be recovered via (\ref{eq:key}). 

Granted, it is very unclear whether any extension of this result might be applicable to higher-dimensional bipartite systems.  Useful extensions seem even more unlikely for mixed states or multi-partite systems, where the relevant features of the Schmidt decomposition are not available.  Nevertheless, the fact of success for two-qubit states does raise the possibility of applying this result in a different direction, to address certain foundational questions.  

One such question is the status of the global phase of a single qubit: is it a meaningless gauge or a potentially interesting hidden variable?  To the extent that manipulation of a local phase $\alpha_1$ or $\alpha_2$ can objectively change the full two-qubit state, these phases certainly cannot treated as a meaningless gauge.  And in the limit of separability, when $\chi\to 0$, these phases do not change in any obvious way: they are still present in the mathematical representation.  (In this limit, they sum to the global phase of the full two-qubit state.)  The above results, then, might encourage us to be more cognizant of single-qubit global phases, as also argued in \cite{WhartonKoch} for symmetry-based reasons.  One might even draw the inference that the two-qubit global phase could be less ignorable than is commonly assumed.

Another interesting foundational question raised by the above results is why a local account of two-qubit states is always available for unitary evolution, but not for measurement.  To review, the required ``local'' parameters for the first qubit are a normalized spinor $\ket{\phi_1}$, and the concurrence angle $\chi$.  Equivalently, one could write these parameters as the Bloch ball vector $\bm{n}^{(1)}$ and the phase angle $\alpha_1$.   (Since $(\bm{n}^{(1)})^2+\mathcal{C}^2=1$, both of these representations effectively utilize a normalized 4-vector and an additional angle.)  A similar local representation is available for the second qubit, with the caveat that the concurrence must be identical.  Since this value does not change under unitary dynamics, one can imagine $\chi$ as the value of two different local parameters, one at the location of each qubit.

One last caveat is in order: these local parameters can encode transformations due to local Hamiltonians of the form $\bm{H_\sigma}$, but not of the form $\bm{H_I}$.  If one wishes to keep track of the phase difference that could be induced by different spin-independent potentials at the locations of each qubit, two more local phases would have to be introduced, $\beta_1$ and $\beta_2$.  These phases would be shifted by  the local value of $\bm{H_I}$.  Combined with the parameters in the previous paragraph, these determine the minimum local parameters that would be required to model the dynamics of an arbitrary two-qubit entangled state (a unit 4-vector and two angles).

Knowledge of these minimal-local parameters might conceivably inform research into the remaining loopholes in Bell's theorem (namely, superluminal and retrocausal influences as a mechanism to explain the distant measurement correlations).  If one wished to develop a fully local hidden variable model, it would need these parameters to account for local unitary transformations.  Given such a framework, it might be interesting to see what precise superluminal or retrocausal influence would be needed to account for the known results of entanglement experiments.

Setting these speculative foundational issues aside, the most important consequence of the above natural parameterization is the simplified method for tracking local unitary transformations on separated-yet-entangled two-qubit states.  Tracking the local phases enables such transformations to be implemented separately on local spinors, rather than on the full state at all times.  The fact that the recurrence angle has not been explicitly identified until now may have been an unfortunate oversight, but hopefully it may now find some use in general analysis of pure two-qubit systems.

\section*{Appendix: Generators of Recurrence}

Consider a local Hamiltonian $\bm{H}$, operating only on the first qubit. In order to generate a rotation that will not change the partial trace, its eigenvalues must be aligned with the partial trace on the Bloch ball.  As this direction is defined by the spherical coordinates $(\theta_1,\phi_1)$ the eigenvectors of $\bm{H}$ must be
\begin{equation}
\label{eq:evs}
\psi_+ = \begin{pmatrix} \cos (\theta_1/2) e^{-i\phi_1/2} \\ \sin (\theta_1/2) e^{+i\phi_1/2} \end{pmatrix} ; 
\psi_- = \begin{pmatrix} \sin (\theta_1/2) e^{-i\phi_1/2} \\ -\cos (\theta_1/2) e^{+i\phi_1/2} \end{pmatrix}.
\end{equation}
Given this local interaction, the entangled state experiences the Hamiltonian $\bm{H} \otimes \bm{I}$.  This is degenerate, having two eigenvectors with the same positive eigenvalue $\Psi_1=\psi_+ \otimes (1\, 0)^T$ and $\Psi_2=\psi_+ \otimes (0\, 1)^T$.  It also has two eigenvectors with the same negative eigenvalue $\Psi_3=\psi_- \otimes (1\, 0)^T$ and  $\Psi_4=\psi_- \otimes (0 \,1)^T$.  

By design, it is simple to decompose (\ref{eq:abcd}) into these four components:
\begin{eqnarray}
\label{eq:proof}
\begin{pmatrix}
 a \\ b \\ c \\ d 
\end{pmatrix}
=& 
\left( \cos \frac{\chi}{2} \cos \frac{\theta_2}{2} e^{-i\phi_2/2} \Psi_1 + \cos \frac{\chi}{2}  \sin \frac{\theta_2}{2} e^{+i\phi_2/2} \Psi_2\right)  e^{+i\gamma/2}+ \\
&\left( \sin \frac{\chi}{2}  \sin \frac{\theta_2}{2} e^{-i\phi_2/2} \Psi_3 - \sin \frac{\chi}{2}  \cos \frac{\theta_2}{2} e^{+i\phi_2/2} \Psi_4\right)  e^{-i\gamma/2}.
\end{eqnarray}

It trivially follows that $\bm{H} \otimes \bm{I}$ is the generator of rotations of $\gamma$; natural time-evolution would simply increase $\gamma$ linearly, as a function of time.  The same argument goes through for $\bm{I} \otimes \bm{H'}$, where $\bm{H'}$ is a local Hamiltonian on the second qubit, aligned with the second qubit's partial trace. 

Interestingly, if both qubits are each locally rotated around their own partial-trace-axis, and the two rotations share the same orientation (say, both right-handed rotations), the change in $\gamma$ compounds rather than cancels -- in agreement with the results of the ``Schmidt evolution'' in \cite{Loredo}.  It is only when both qubits are rotated in opposite orientations that $\gamma$ can remain constant.  

\section*{Acknowledgements}
The author is very grateful to Jerome Finkelstein for crucial insights and to Yun Xuan Shi for supporting calculations.

\end{document}